\newif\ifsubmode
\submodefalse

\newif\ifprintfig
\printfigtrue

\newif\ifemulate
\emulatetrue

\ifsubmode
  \documentstyle[12pt,aasms4,epsf]{article}
  \received{}
  \accepted{}
  \journalid{}{}
  \articleid{}{}
\else
   \ifemulate
     \documentstyle[emulateapj,epsf,graphicx]{article}
     \submitted{{\it Accepted for publication in AJ}}
   \else
     \documentstyle[11pt,aaspp4,epsf]{article} 
     \slugcomment{{\it Accepted for publication in AJ}}
   \fi
\fi

\lefthead{Willman et al}
\righthead{Survey for Resolved LG Dwarf Galaxies}

\def\lesssim{\mathrel{\hbox{\rlap{\hbox{\lower4pt\hbox{$\sim$}}}\hbox{$<$}}}}
\def\gtrsim{\mathrel{\hbox{\rlap{\hbox{\lower4pt\hbox{$\sim$}}}\hbox{$>$}}}}

\begin{document}

\title{An SDSS Survey For Resolved Milky Way Satellite Galaxies I: Detection Limits}

\author{Beth Willman\altaffilmark{\ref{Washington}}, Julianne Dalcanton\altaffilmark{\ref{Washington}}, \v{Z}eljko Ivezi\'{c}\altaffilmark{\ref{Princeton}}, Tom Jackson\altaffilmark{\ref{Princeton}}, Robert Lupton\altaffilmark{\ref{Princeton}}, Jon Brinkmann\altaffilmark{\ref{APO}}, Greg Henessy\altaffilmark{\ref{USNO}}, Robert Hindsley\altaffilmark{\ref{RSD}}  }

%%%%%%%%%%%%%%
% Additional affiliations
%%%%%%%%%%%%%%

\newcounter{address}
\setcounter{address}{1}
\altaffiltext{\theaddress}{University of Washington, Department of Astronomy,
Box 351580, Seattle, WA 98195
\label{Washington}}

\addtocounter{address}{1}
\altaffiltext{\theaddress}{Princeton University Observatory, Princeton, NJ 08544
\label{Princeton}}

%\addtocounter{address}{1}
%\altaffiltext{\theaddress}{University of Chicago, Astronomy \& Astrophysics
%Center, 5640 S. Ellis Ave., Chicago, IL 60637
%\label{Chicago}}

\addtocounter{address}{1}
\altaffiltext{\theaddress}{Apache Point Observatory, P.O. Box 59, Sunspot, NM 88349-0059
\label{APO}}

\addtocounter{address}{1}
\altaffiltext{\theaddress}{United States Naval Observatory, 3450 Massachusetts Ave., NW, Washington DC 20392-5420
\label{USNO}}

\addtocounter{address}{1}
\altaffiltext{\theaddress}{Remote Sensing Division, Code 7215, Naval Research Laboratory, 4555 Overlook Ave., SW, Washington DC 20375
\label{RSD}}

%%%%%%%%%%%%%%%
% Start the abstract on a fresh page
%%%%%%%%%%%%%%%

\ifsubmode\else
  \ifemulate\else
     \clearpage
  \fi
\fi

%%%%%%%%%%%%%%%
% Use a small baselineskip, unless in submission mode.
%%%%%%%%%%%%%%%

\ifsubmode\else
  \ifemulate\else
     \baselineskip=14pt
  \fi
\fi

%%%%%%%%%%%%%%%
% Abstract
%%%%%%%%%%%%%%%
\begin{abstract}

We present the detection limits of a new survey for resolved low
surface brightness satellite galaxies to the Milky Way, based on the Sloan Digital Sky Survey (SDSS).  Our survey exploits SDSS's major strengths
(multi-color photometry, depth, large-scale, and uniformity) by combining filter smoothing with limits in both magnitude and color space to search for low surface brightness galaxies and stellar counterparts to the Compact High Velocity Clouds out to the Milky Way's virial radius ($\sim 350$ kpc).  Our calculated detection limits for a purely old stellar population vary with galaxy
size and distance between $~\mu_{V,0} =$ 26.7 and 30.1 mag/$\Box ''$. These limits will allow us to detect systems whose surface brightnesses are .5-3.9 mag/$\Box ''$ fainter than Sextans, the lowest surface
brightness Local Group member known.  Our survey not only is sensitive to lower surface brightness stellar populations
than possible with previous Local Group surveys, but will also allow us to make an unbiased and well defined assessment of the completeness of the
observed Local Group galaxy luminosity function, so that we may compare the results with the predictions of various structure formation scenarios.

\end{abstract}

%%%%%%%%%%%%%%%
% Keywords
%%%%%%%%%%%%%%%

\ifemulate\else
   \clearpage
\fi

%%%%%%%%%%%%%%%
% Beginning of main text
%%%%%%%%%%%%%%%

\section{Introduction}
\label{sec:intro}
Detailed investigation of the faint end of the Local Group (LG) luminosity function
yields several constraints on the process of galaxy formation.
First, the faint end of the LG luminosity function hosts the highest
M/L, lowest mass and lowest luminosity systems accessible for study,
allowing us to probe the extreme limits of galaxy formation pathways.
Second, these galaxies are sufficiently close that they can be
resolved into individual stars, permitting direct study of their
detailed star formation history and their chemical and dynamical evolution.
Third, the number of the lowest mass LG halos and their observed mass profiles can be used to help distinguish between various dark matter,
structure formation, and reionization scenarios.

In particular, results of recent N-body simulations have reignited a
need for a well-defined and unbiased measurement of the galaxy
mass function down to the lowest possible scales.  These simulations suggest
that both "standard" and $\Lambda$+Cold Dark Matter (CDM) cosmologies overpredict the number of low mass
dark matter halos around the Milky Way and M31 by over an order of
magnitude (Klypin et al. 1999, Moore et al. 1999), compared
to existing catalogs of the Local Group (see \S2).  Many solutions to this
discrepancy between theory and observations have been suggested,
including: modification of the traditional CDM paradigm with warm
(Hogan \& Dalcanton 2000) or self interacting dark matter (Spergel \& Steinhardt
2000) and inclusion of dynamical friction, tidal stripping and
reionization on the results of cosmological formation simulations (Bullock, Kravtsov \& Weinberg 2000a and b).  However, the possibility remains that the
discrepancy at least partially results from under counting the population of dwarfs in the Local Group (e.g. the predictions of Benson et al. 2001).

Roughly 40 dwarf galaxy members of the Local Group have
been identified to date, 9 of which are dwarf spheroidal companions to the
Milky Way (Mateo 1998).  Several of these LG dwarfs have only been discovered within the last few years.  Empirically, these recent discoveries of
new Local Group members (for example, Whiting, Hau \& Irwin 1999,
Armandroff, Davies \& Jacoby 1998, Karachentsev \& Karachentseva 1999) suggest that we have not necessarily identified all of the
faint Local Group galaxies accessible with existing large-sky survey
data.

These recent discoveries resulted from large-area searches that attempted to rectify incompleteness in Local Group membership.  The surveys have generally fallen into one of
two categories: $i)$ visual searches of unprocessed POSS-II (Karachentsev \& Karachentseva 1999) or ESO/SERC survey plates (Whiting, Irwin \& Hau, 1997, Whiting,
Hau \& Irwin 1999, Karachentsev et al. 2000, Karachentseva \& Karachentsev 2000) or $ii)$ visual searches of digitally scanned and filtered POSS-II images (Armandroff, Davies \& Jacoby 1998, Armandroff, Jacoby \& Davies, 1999).  Even these most recent identifications of new Local Group
members have, on some level, required visual evaluation of images, and
thus are prone to selection effects which are difficult to quantify.
They are also fundamentally limited in surface brightness, due to 
the limited contrast of sparse dwarf spheroidals against the noise
of the sky background.

We anticipate that substantial progress towards completeness and uniformity in the known luminosity function can be made with the Sloan
Digital Sky Survey (SDSS) (York et al 2000), an on going 5-color large area CCD survey that
will encompass a quarter of the celestial sphere.  Instead of
searching for smooth overdensities of diffuse stellar light, we will isolate
overdensities of resolved stars, taking advantage of SDSS's faint photometric limits, multiple filters, and
robust star-galaxy separation (Ivezi\'{c} et al 2000).

A survey based on resolved stars has a number of benefits.  
Use of resolved stars allows us to substantially
suppress galactic foreground stars relative to a galaxy's resolved stellar
population (see \S3 and 5) by analyzing data in small bins of color
and/or magnitude.  As we show below, these cuts can greatly increase
the contrast of the dwarf galaxy against the background, and should
allow us to reach unprecedentedly low surface brightnesses.  Also, unlike surveys for Local Group galaxies
based on diffuse light, a survey using a red color cut and resolved
stellar counts will not be biased towards detecting galaxies with
recent star formation. By identifying galaxies by their resolved stars, we remove a bias towards detecting galaxies with
young main sequence stars, which contribute a disproportionate fraction of the total light of a galaxy relative to the number of stars they represent.

In this paper, we describe the particular survey we have undertaken
with the SDSS imaging data set.  We begin by describing the likely
prospects for the existence of undetected dwarf galaxies in
\S\ref{sec:Nbody}.  We then describe the actual survey in
\S\ref{sec:survey}, calculate the detection limits in
\S\ref{sec:detectioncalc}, present the results in
\S\ref{sec:limits}, and demonstrate our approach using the Draco dSph in \S\ref{sec:draco}.  The prospects for future work on the survey and
the application to the regions near Compact High Velocity
Clouds (Braun \& Burton 1999, see also Willman et al, in
preparation) are then considered in \S\ref{sec:conclusion}. 

\section{Predictions of Existing N-body Simulations}	\label{sec:Nbody}

The first issue we wish to investigate is whether there is any compelling
expectation that new Local Group dwarfs might remain undetected
within the virial radius of the Milky Way.  If CDM is the correct
description of dark matter, then the current generation of N-body
simulations suggest that there is indeed fertile ground for
discovering new dwarfs in our local neighborhood.  For example,
\cite{moore99}'s CDM simulations predict over 400 substructures within the Milky Way's virial radius, roughly 10 of which have a circular velocity, $v_c \gtrsim$ 30 km/sec (where $v_c \equiv (Gm_{bound}/r_{bound})^{1/2}$).  Assuming a homogeneous spatial distribution, these numbers
translate to approximately 100 subhalos within the volume that we can
probe with SDSS.  For halos with $v_c \lesssim$ 30 km/sec, these
numbers are significantly reduced by reionization, dynamical
friction, tidal stripping and gas dynamics (Thoul \&  Weinberg 1996, Quinn et al. 1996, Bullock, Kravtsov, \& Weinberg 2000a and b, Benson et al. 2001), although the
degree to which these processes play a role is uncertain.  For example, the results of recent simulations of the tidal stirring of dwarf galaxies (Mayer et al 2001) suggest that previous estimates of tidal disruption as a means to destroy low mass halos may have been overestimated. 

The above modifications of CDM simulations may well produce observable consequences. For example, if the central densities of dark matter halos increase with decreasing mass, as CDM predicts,  then there may be observable dense stellar cores left as remnants of any disruption process.   Likewise, if CDM is not the correct form
of dark matter, then the number of halos is also likely to be lower.

Independent of the number of halos which may be expected in a
particular cosmological scenario, our survey will be
sensitive to a range of galaxy structural parameters which have been
previously inaccessible to existing surveys.  We anticipate that we
could detect new galaxies $i)$ with comparable masses and distances of
known MW companions but higher mass-to-light ratios, and thus lower
surface brightnesses and/or $ii)$ those galaxies with similar
mass-to-light ratios as known MW companions but lower masses, and thus
lower surface brightnesses.  It is unclear whether such lower mass
systems would host a detectable stellar population, as any gas
initially accreted in such halos may be unable to cool, and could
``boil out'' after reionization (see Barkana \& Loeb 1999).  However, the detection or non-detection of lower mass
dwarfs will place constraints on any such line of reasoning.

\section{Methodology of Survey}			\label{sec:survey}

\subsection{Sloan Digital Sky Survey Data}

The Sloan Digital Sky Survey will cover $\pi$ steradians in the North
Galactic Cap.  The sky will be observed nearly simultaneously through
5 broad photometric filters ($u', g', r', i'$ and $z'$) and will be
complete (S:N $\sim$ 5:1) to limiting magnitudes of roughly 22.3,
23.3, 23.1, 22.3 and 20.8 respectively (see \cite{f96} for details of
Sloan's photometric system).  The photometric calibration system for
SDSS filters has not yet been finalized, so the preliminary magnitudes
as presented in this paper will be denoted as $u^*$,$g^*$,$r^*$,$i^*$,
and $z^*$.  The system used in this paper will differ absolutely from
the final photometric system by only a few percent in every band
except for $u'$, where it will differ by no more than $10\%$ (Yanny et
al 2000).  The imaging data are taken in drift-scan mode by a camera
consisting of 30 2048x2048 CCDs (see \cite{g98} for details on the
camera used for SDSS).  SDSS's accurate astrometry combined with color
and morphological information derived from the digital images provides
robust star-galaxy separation to a limiting magnitude of 21.5 (Ivezi\'{c}
et al 2000).  See \cite{york00} and references therein
for an overview of the project and further details about the
photometric survey.

For this paper, we used data from runs 94 and 125 taken in Spring '99 and Fall '98.  Each run
is composed of 6 parallel strips that are separated by 0.2 degrees.
The full 2.5 degree wide equatorial stripe is then composed of 12
interleaved strips from 2 slightly offset runs observed a few
days apart (Yanny et al 2000).

% The interleaving of two runs should produce a seamless 2.5 degree wide
%stripe where the data taken in 2 different runs are indistinguishable
%from each other.  Consistent star-galaxy separation between the 2 runs
%that compose a full stripe is vital for a homogeneous stellar source
%distribution.  A stellar classification in SDSS requires $psf$
%$magnitude - model$ $magnitude < 0.2$ (define psf and model mag??) in
%one or more bands.  When using this test data taken during the commissioning phase of SDSS, we require a source to be classified as a star in
%both the $g^*$ and $r^*$ bands for it to be used in our analysis.  In
%cases were this criterion produces stellar counts as a function of
%$r^*$ that differ significantly between the two runs, we decrease the
%$psf$ $magnitude - model$ $magnitude$ threshold in $r^*$ in the
%overpopulated run until the counts in the two runs are commensurate
%within two Poisson $\sigma$.

All magnitudes presented in this paper have been corrected for
reddening due to Galactic extinction with the maps of
\cite{schlegel98}.

\subsection{Survey Technique}

We identify candidate dwarf galaxies as spatial overdensities of resolved stars in subfields of SDSS data.  Assuming foreground stars are approximately spatially uncorrelated, Poissonian fluctuations govern the random noise in the stellar
foreground. Therefore, to increase the contrast of extragalactic systems against the foreground of Milky Way stars, we apply $r^*$ magnitude binning and color selection criteria to the stellar sources in a field.  Stars which pass the criteria are binned into a spatial array, which is then smoothed with an exponential filter.  Candidates are automatically identified as overdensities in the smoothed sky map.  Each field of sky is analyzed multiple times, once for each $r^*$ magnitude bin combined with either a red color cut or no color cut.
 
These techniques have been utilized in different forms in other
Local Group surveys.  For example, Armandroff et al (1998) convolved
digitized red POSSII images with a "square spatial median filter" before
visually identifying candidates for unresolved M31 companions.  Other
groups working with SDSS data have used a number of similar methods to trace substructure in the Milky Way's halo.  Yanny et al. (2000) demonstrate that a color cut can isolate faint A stars, which can be used to trace nearby halo substructures when analyzed in position-magnitude space.  Odenkirchen et al. (2001a and b) use specialized color-magnitude templates tailored for the stellar populations of Palomar 5 and Draco, to trace their structure to fainter levels than possible with the generalized scheme presented here.

\subsubsection{Color Cut for Red Stellar Sources} In galaxies more distant than 125 kpc, only red giant branch and (for the nearest systems) horizontal branch stars are bright enough to be resolved by SDSS.  Therefore, to optimize analyses for extragalactic systems, we have chosen a color-color cut to select stars with colors consistent with those of a metal poor red giant branch population (although Sloan's photometric system may not be able to distinguish a red giant from a red dwarf).
A cut in the colors $g^*-r^*$ and $r^*-i^*$ (Figure~\ref{fig:fieldcut}) was selected based upon Krisciunas et al's (1998)
observation of the globular cluster M15 ([Fe/H] = -2.22) through Sloan
 filters, Lenz et al's (1998) theoretical colors of stars in Sloan
filters for a range of metallicities and surface gravities, and a linear fit to the observed SDSS stellar locus (Ivezi\'{c}, private communication) coupled with a color-magnitude diagram (CMD) of the globular cluster Palomar 5 ([Fe/H] = -1.38) created with SDSS commissioning data (Figure~\ref{fig:GCcmd}).  Because the average photometric errors in the faintest magnitude bins are $\ge$ the half width of the color cut in $r^*-i^*$ (0.05 mag), stars are only required to have a $25\%$ chance of fulfilling the color criteria, given photometric errors, to satisfy the cut.  We chose this lower probability limit to be less than $50\%$ so that, when photometric errors are considered,  stars whose quoted colors lie outside the color cut may scatter back in.  In particular, we chose the limit to be no less than $25 \%$ so that only stars whose colors are less than $1\sigma$ outside of the color cut may scatter back in.   This loose restriction on the color cut also allows for the fact that the psf magnitude $errors$ quoted in the survey are systematically underestimated by $10-20\%$ (Stoughton et al., 2001). 

\placefigure{fig:fieldcut}
\placefigure{fig:GCcmd}
\placefigure{fig:GCcut}
When the red color cut is applied to
field stellar sources with 17.0 $<r^* <$ 21.5, as shown in
Figure~\ref{fig:fieldcut}, approximately 3/4 of field stars are
eliminated.  Figure~\ref{fig:GCcut} shows the color cut applied to the
Galactic globular cluster Palomar 5, where the cluster stars that satisfy the cut are denoted with $\diamondsuit$s.  In Figure~\ref{fig:GCcut}, the red
giant branch colors would shift down and to the left with decreasing metallicity.  

In \S5 we demonstrate that we expect this color cut to enhance detection sensitivity for galaxies more distant than $\sim$ 50 kpc.  Closer than this, however, main sequence stars, that do not satisfy the color cut, make a substantial contribution to the resolved stellar luminosity function (LF), and including the cut reduces our sensitivity.  Thus, we chose to analyze the data both with and without the red cut, to optimize our sensitivity to both distant and nearby galaxies, respectively.

\subsubsection{Binning in Magnitude Space}  We can also increase our sensitivity by limiting our analysis to small bins in magnitude space.  Limiting our analysis by magnitude serves to isolate the subset of the resolved stellar population that produces the strongest signal over the foreground noise.  For example, the only stars resolved in the most distant old population systems within our virial radius would be red giant branch stars with $r^* >$ 20.0.  Including stellar sources brighter than $r^* =$ 20.0 in the analysis adds greatly to the noise and nothing to the signal.  Likewise, for a somewhat closer system, the magnitude bin containing a more densely populated portion of the lower red giant branch would provide the strongest signal over the foreground noise.

We apply the magnitude cuts in overlapping bins of 1.5 magnitude width spanning $17.0 < r^* < 21.5$.  The upper limit of this range represents the magnitude to which the star-galaxy separation is reliable (eg. at $r^*$ = 21.5 the star-galaxy classification is repeatable at a level better than $90\%$) (Lupton et al. 2001, in preparation, and Yasuda et al 2001). 

\subsubsection{Smoothing with Exponential Filter}
Many groups have found that spatial smoothing increases their sensitivity to LSB galaxies (for example, \cite{davies94} and \cite{adj98}).  A smoothing filter serves to suppress random fluctuations in foreground field stars, increasing the signal-to-noise of fluctuations on scales comparable to the size of the smoothing kernel.  \cite{davies94} show that convolution filters that match the
scale length of a low surface brightness (LSB) galaxy within a factor of 2 still serve to nearly maximally enhance the galaxy signal for detection. For our general survey, we will smooth the spatial array of stellar sources that
satisfy color and magnitude cuts with a $3'$ exponential filter, to optimally enhance the signal from galaxies with physical scale
lengths matching those of the known LG dwarfs and with
distances between 30 and 350 kpc.  We will also use a 7' smoothing
filter in a more targeted analysis of regions containing known Compact High Velocity Clouds, to better match their observed sizes (see Braun \& Burton 1999).

\subsection{Detection Criteria}

Once the stellar sources have been restricted in magnitude and color, and their surface density spatially smoothed, we use a minimum signal threshold and a minimum size as the two primary criteria for candidate selection. 

% The detection criteria have been chosen to minimize the number
%of false detections expected, while sacrificing the least number of
%possible true detections.

The minimum signal threshold is defined as a surface density fluctuation which is $5\sigma$ above the stellar foreground density. We choose the signal threshold such
that, for a given smoothing length, there is no more than $~1$ expected false detection per 100 square degrees of sky, in the absence of any other discriminant besides minimum size.  Contiguous regions whose signal exceeds the threshold and are larger than $\pi\times$(smoothing length/2)$^2$, satisfy the minimum size requirement. This requirement eliminates many spurious detections of only one or two square arcminutes.  

%The number of
%independent regions in a given surface area of the sky increases as
%the smoothing length of the filter decreases, therefore we chose a higher threshold for our smaller smoothing length.

We create a color-magnitude diagram each galaxy candidate's resolved stars, which we can use to identify the source of the detection and discriminate against possible contaminants.  Possible contaminants will include both completely false
detections due to rare random positive fluctuations and clusters of galaxies, which may be a problem in the faintest
magnitude bins where star-galaxy separation is difficult.

\section{Determining Detection Limits}	\label{sec:detectioncalc}
The detection limits of this survey are primarily a function of $i)$ the stellar surface densities produced by different types of potential new Local Group galaxies and $ii)$
the fluctuation level of the foreground stellar surface density above which
galaxies must be detected. We simulated the stellar surface density of old population Local Group galaxies at a range of sizes and distances, by creating template stellar luminosity functions from SDSS observations of the globular cluster Palomar 5.  We used a foreground stellar field from the SDSS data to determine the effect of color and magnitude criteria on the foreground fluctuation level.

\subsection{Stellar Luminosity Function of Palomar 5} 
Located 22.6 kpc
from the Sun, Palomar 5 (Pal 5) has a large core radius ($r_{core}$ = 2.9 pc), a
low central surface brightness ($\mu_{V,0} = 24.67$) and a moderately
high metallicity ([Fe/H] = -1.38) for a globular cluster (Harris 1996).
\cite{s86} use isochrone fitting to infer an age of 14-15 Gyr for the cluster, which is comparable to the ages of the oldest populations in most
known LG dwarf galaxies (Mateo 1998).  The metallicities of the known Local Group dwarfs range from [Fe/H] = -0.8 to -2.2, which are also comparable to the metallicity of Palomar 5.

\placefigure{fig:GClf}

We note that the stellar population of Palomar 5 may contain departures from that of a "typical" globular cluster (GC).  Low mass stars and stars in the base of the subgiant branch appear to
be underpopulated (Smith et al 1996) and the horizontal branch appears
to be overpopulated relative to the main sequence (Sandage \&
Hartwick 1977) when compared with the luminosity function of M3.  These departures are likely due to the ongoing tidal disruption of the Pal 5 system (Odenkirchen et al. 2001a).  Despite these possible departures from the typical stellar distribution of an old population, Pal 5's proximity and low central surface brightness allow us both to $i)$ characterize an old stellar population's CMD beyond the main sequence turnoff and $ii)$ not worry about the effects of crowding in the central regions when creating the stellar templates from the data.  In our final analysis of the survey results when the SDSS is complete, we will be able to use a superposition of a number of observed dwarf galaxy populations to create a more precise template.  However, we have already repeated the entire analysis presented in this paper with the SDSS observations of the Draco dSph (see \S 6), and the resultant detection limits differed from the ones determined using only Pal 5 as the template, from insignificantly up to a couple tenths of a magnitude (depending on the galaxy size and distance).

We constructed Pal 5's stellar LF from the SDSS data by magnitude binning its stars, with and without a color cut, and subtracting the mean stellar luminosity function of the surrounding field.  Figure~\ref{fig:GClf} shows the resulting luminosity functions.  We then projected the LF to systems with distances ranging from $d_{Pal5}$ to roughly 350 kpc.  The new LF was constructed by shifting the color-magnitude diagram fainter and recalculating the luminosity function. The surface density of stellar counts was then scaled by distance squared. 

\subsection{Calculating Limiting $\mu_{V,0}$}

We calculate the survey's limiting $\mu_{V,0}$ for galaxies with a range of exponential scale lengths and distances. To determine the limits, we created numerous galaxies with a range of surface brightnesses, sizes, and distances, placed them in constructed spatial arrays of foreground stars and applied our survey algorithm to the resultant artificial data sets.  

Our simulated galaxies were given exponential scale lengths and distances that ranged between 1$'$ and 13$'$ and 23 to 350 kpc.  For each set of parameters, we created 500 trial galaxies in a range of surface brightness centered around an initial 'best guess' detection limit.  Each trial galaxy was populated with an integral number of stars drawn from a Poisson distribution of the Palomar 5 stellar surface density template appropriate to the distance and scaled to the appropriate surface brightness for the specific trial.

We created the stellar foreground spatial array by calculating the surface density of stellar sources
observed by SDSS around $\ell \sim 96^{\circ}, b \sim -60^{\circ}$ and assuming a random stellar distribution. This assumption will tend to underestimate $\sigma$, since foreground stars may be somewhat correlated.  

Once we analyzed the artifical data sets with our survey algorithm, we did a linear fit to determine the surface brightness at which there is a 50$\%$ detection efficiency for galaxies at each combination of size and distance.

\subsection{Caveats}

There are a few important caveats which can affect the universality of the detection limits calculated above.  First, all of the limits presented are dependent on the details of a stellar system's star
formation history.  The fact that Palomar 5 is a purely old population
stellar system results in detection limits for only the old stellar
population in a galaxy.  Recent star formation adds more to a
galaxy's surface brightness than to the overall stellar counts, so our
calculated $\mu_{V,0}$'s are only upper limits (lower limits in brightness) for galaxies with star formation within the last 10 Gyr.

Second, the limits are dependent on the particular $\ell$ and $b$ of the foreground field.  We calculated detection limits for a direction representative of the region $b > 30^{\circ}$, but the actual limits will be brighter as one looks closer to the Galactic plane and, likewise, fainter as one looks towards the Galactic pole. 

Also, as mentioned above, we may be underestimating $\sigma$ of the stellar foreground due to our assumption of a random foreground stellar distribution.

\section{Resulting Detection Limits}			\label{sec:limits}
%\subsection{Limiting Central Surface Brightness}

\placefigure{fig:filtersmulin}

Figure 5 displays the detection limits of our survey, performed both with and without the red color cut, as a function of galaxy physical scale length and distance.  Each line follows the detection limit for a galaxy of a given exponential scale length (in pc) as it moves further away.  Open points represent the known Milky Way galaxy companions fainter than $\mu_{V,0} = 25.0$, including Sextans, the lowest surface brightness Local Group member with a $\mu_{V,0}$ of 26.2 (Mateo 1998).  These points are given as lower limits, since the limiting $\mu_{V,0}$ is calculated for a purely old stellar population (see \S 4.3), and some dwarf spheroidals show signs of more recent star formation. Comparison of the two panels shows that for all but the closest distances, the red color cut makes a notable improvement in galaxy detectability.

The most obvious and general result of our analysis is that our survey will be sensitive to dwarf galaxies, within the Galaxy's virial radius, with unprecedentedly low surface brightnesses.  In addition to this, our limits display 2 general trends: 1) larger scale length systems are easier to detect at a given distance and 2) the red color cut causes systems with $h_R\gtrsim  250$ pc to have a SNR that does not decrease with distance.  The first trend simply results from the fact that the signal from a galaxy with a larger angular scale length falls off less quickly than that with a smaller length.  The second trend, although perhaps initially counterintuitive, is due to two combined effects: a system's surface brightness remains constant with distance, but its stellar density increases with distance squared, and outer halo stellar populations can only be detected in the magnitude bins containing their RGB stars, which are specifically selected by the color cut.

\section{Application of Survey Technique to Draco dSph}  \label{sec:draco}
To demonstrate our methodology and to test the feasibility of the detection limits presented above, we analyzed a field centered on the Draco dSph galaxy, which was recently by SDSS in data runs 1356 and 1359.  Draco has a distance of 82 kpc, $r_{exp}$ of 4.5$'$, and a central surface brightness $\mu_V = 25.3$ mag/$\Box ''$.  Draco is one of the two least luminous galaxy in the Local Group, with $M_V = -8.8$ and $M_B = -7.8$ (Mateo 1998).
\placefigure{fig:dracoimage}
Figure 6 shows a smoothed stellar density map of Draco with overlaid contours, where each contour represents a $10\sigma$ fluctuation above the mean smoothed stellar distribution.  This map was constructed from stars with 19.5 $\leq r^* \leq$ 21.0 that pass the red color cut.  One can see that our survey identifies the central regions of Draco at a level of $50\sigma$ above the field.

This result confirms that on the angular scale of Draco, we should be able to detect systems at least 2.5log($\frac{50\sigma}{5\sigma}$) mag/$\Box ''$ fainter than Draco, or $~27.8$mag/$\Box ''$.  This is in good agreement with our expectations from Figure 5 (roughly $~28.6$mag/$\Box ''$ for a $4.5'$ scale length galaxy at 80 kpc).  The .8 magnitude offset between the Draco observation and our calculated limit is easily accounted for by the facts that $i)$ Draco lies $30^\circ$ closer to the galactic plane than the region we used to calculate our detection limits, so the density of the stellar foreground is about $1.75$x greater than our artificial foreground field and $ii)$ we have assumed a spatially uncorrelated foreground distribution in our calculations.

\section{Conclusion}		\label{sec:conclusion}

We have presented approximate detection limits of a new Local
Group survey for resolved low surface brightness galaxies.  Our
detection limits for a purely old stellar population vary with galaxy
size and distance between about $\mu_{V,0} =$ 26.7 and 30.1, and are
.5-3.9 mags/$\Box ''$ fainter than the lowest surface brightness Local Group members known.

Our survey will provide a much needed systematic analysis of the content of the Local Group for comparison with theoretical cosmological
models of structure formation.  Independent of whether or not we do detect additional faint end Local Group galaxies or stellar
counterparts with the Compact High Velocity Clouds, our calculated
detection limits will quantitatively describe the region of unsampled
Local Group parameter space for the first time and, therefore, still
yield valuable constraints on various structure formation scenarios.

%%%%%%%%%%%%%%%
% Acknowledgments
%%%%%%%%%%%%%%%

\acknowledgements

This work was supported at the University of Washington in part by NSF grant AST 99-0862 and in part by NSF grant AST-0098557.  We would also like to thank the anonymous referee for helpful suggestions.

The Sloan Digital Sky Survey (SDSS) is a joint project of The
University of Chicago, Fermilab, the Institute for Advanced Study, the
Japan Participation Group, The Johns Hopkins University, the
Max-Planck-Institute for Astronomy (MPIA), the Max-Planck-Institute for Astrophysics (MPA), New Mexico State University,
Princeton University, the United States Naval Observatory, and the
University of Washington.  Apache Point Observatory, site of the SDSS
telescopes, is operated by the Astrophysical Research Consortium
(ARC).

Funding for the project has been provided by the Alfred P. Sloan
Foundation, the SDSS member institutions, the National Aeronautics and
Space Administration, the National Science Foundation, the
U.S. Department of Energy, the Japanese Monbukagakusho, and the Max Planck Society.  The
SDSS Web site is http://www.sdss.org/.

%%%%%%%%%%%%%%%
% Use a small baselineskip for the references, unless in submission mode.
%%%%%%%%%%%%%%%

\ifsubmode\else
\baselineskip=10pt
\fi

%%%%%%%%%%%%%%%
% Reference List
%%%%%%%%%%%%%%%

\clearpage

%%%%%%%%%%%%%%%
% Change back to the regular baselineskip, if necessary
%%%%%%%%%%%%%%%

\ifsubmode\else
\baselineskip=14pt
\fi

%%%%%%%%%%%%%%%
% Figure Captions
%%%%%%%%%%%%%%%

\newcommand{\figcapfieldcut}{Color-color diagram of stellar sources, $15.0 < r^* < 21.5$, in SDSS commissioning data.  Polygon outlines the region included in the red color cut.\label{fig:fieldcut}}

\newcommand{\figcapGCcut}{The red color cut applied to color-color diagram of stellar sources in the Galactic globular cluster Palomar 5  ([$\frac{Fe}{H}$] =  -1.35) without background subtracted. $\diamondsuit$ denote the cluster stars that satisfy the color cut. \label{fig:GCcut}}

\newcommand{\figcapGCcmd}{Color-magnitude diagram of Palomar 5's core ([$\frac{Fe}{H}$] =  -1.35) as observed by SDSS.  No field subtraction has been performed. $\diamondsuit$ denote the stars that satisfy the red color cut. \label{fig:GCcmd}}

\newcommand{\figcapGClf}{Stellar luminosity function of Galactic globular cluster Palomar 5 as observed in SDSS, with and without the red color cut.  The solid line outlines the LF of all Palomar 5 stars and the dashed line outlines the LF of Palomar 5 stars that satisfy the red color cut . \label{fig:GClf}}

\newcommand{\figcapfiltersmulin}{Comparison of central surface brightness limits for our survey conducted with a 3' smoothing filter without and with the red color cut, respectively.  Each line represents systems of different physical scale lengths and the filled points represent the angular scale lengths that correspond to the plotted combinations of distance and physical scale lengths, as summarized in the legend.  Open points represent the known Milky Way galaxy companions fainter than $\mu_{V,0} = 25.0$.\label{fig:filtersmulin}}

\newcommand{\figcapdracoimage}{Smoothed stellar image of Draco dSph.  Each contour represents a $10\sigma$ fluctuation above the mean smoothed stellar density.  The center of Draco is detected at $50\sigma$.\label{fig:dracoimage}}

%%%%%%%%%%%%%%%
% Print Figures (Only if printfigtrue)
%%%%%%%%%%%%%%%

\ifsubmode
\figcaption{\figcapfieldcut}
\figcaption{\figcapGCcmd}
\figcaption{\figcapGCcut}
\figcaption{\figcapGClf}
%\figcaption{\figcapfiltersmunocutang}
\figcaption{\figcapfiltersmulin}
\figcaption{\figcapdracoimage}
\clearpage
\else\printfigtrue\fi

\ifprintfig

\clearpage
\begin{figure}
\epsfxsize=14.0truecm
\centerline{\epsfbox{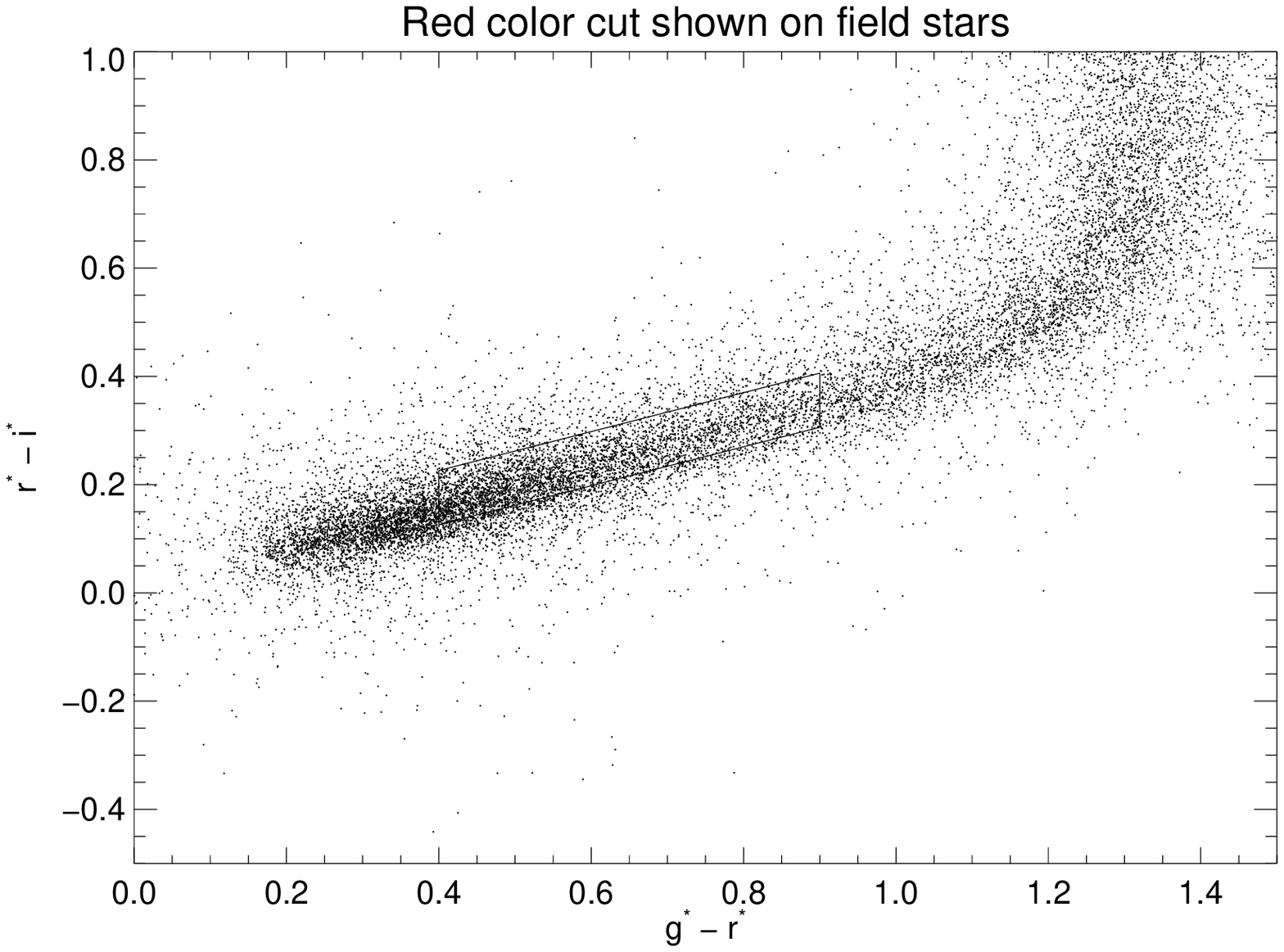}}
%\centerline{\epsfbox{Willman.fig1.eps}}
\ifsubmode
\vskip3.0truecm
\addtocounter{figure}{1}
\centerline{Figure~\thefigure}
\else\figcaption{\figcapfieldcut}\fi
\end{figure}
\clearpage
\begin{figure}
\epsfxsize=14.0truecm
\centerline{\epsfbox{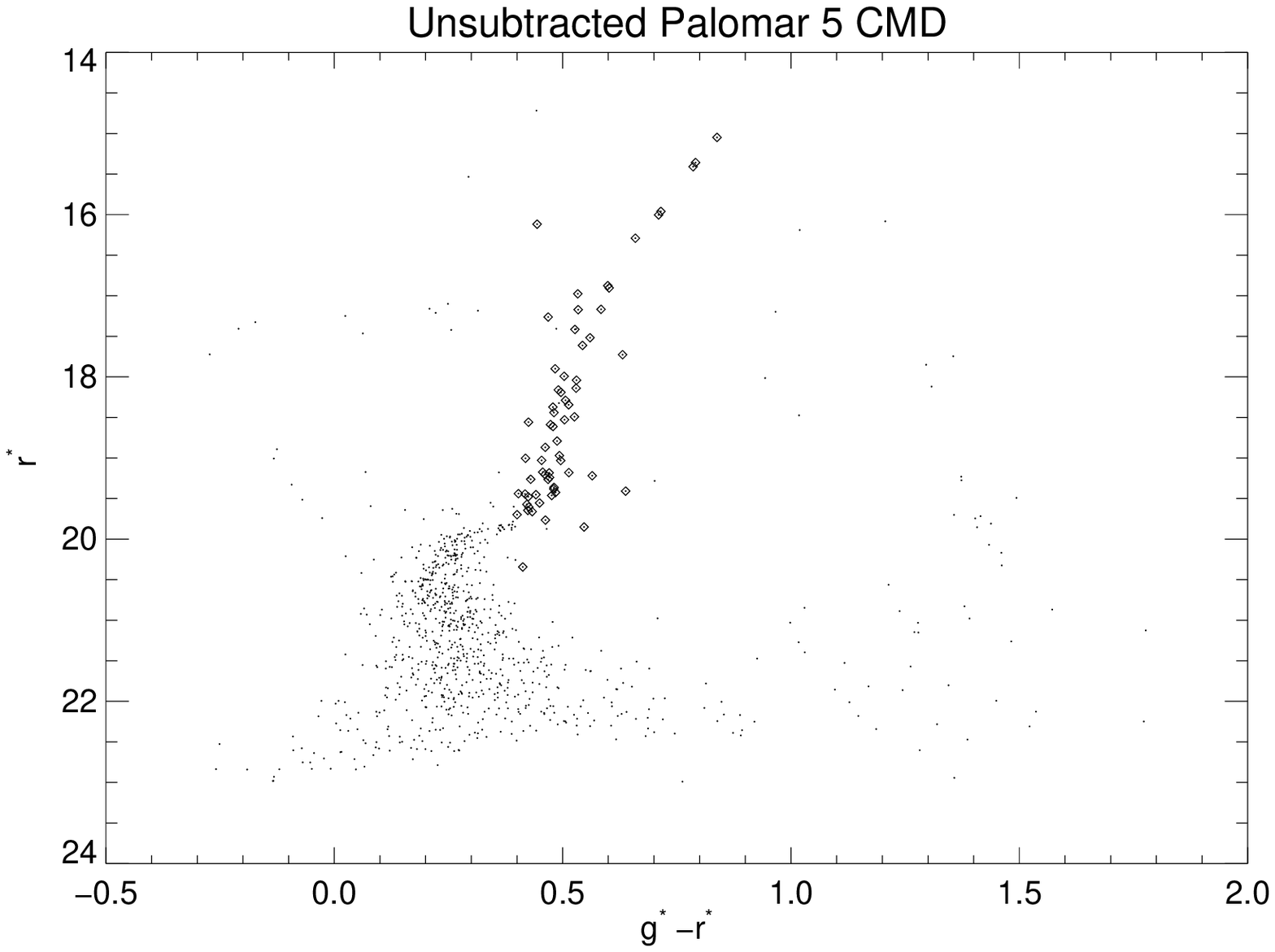}}
%\centerline{\epsfbox{Willman.fig2.eps}}
\ifsubmode
\vskip3.0truecm
\setcounter{figure}{0}
\addtocounter{figure}{1}
\centerline{Figure~\thefigure}
\else\figcaption{\figcapGCcmd}\fi
\end{figure}
\clearpage
\begin{figure}
\epsfxsize=14.0truecm
\centerline{\epsfbox{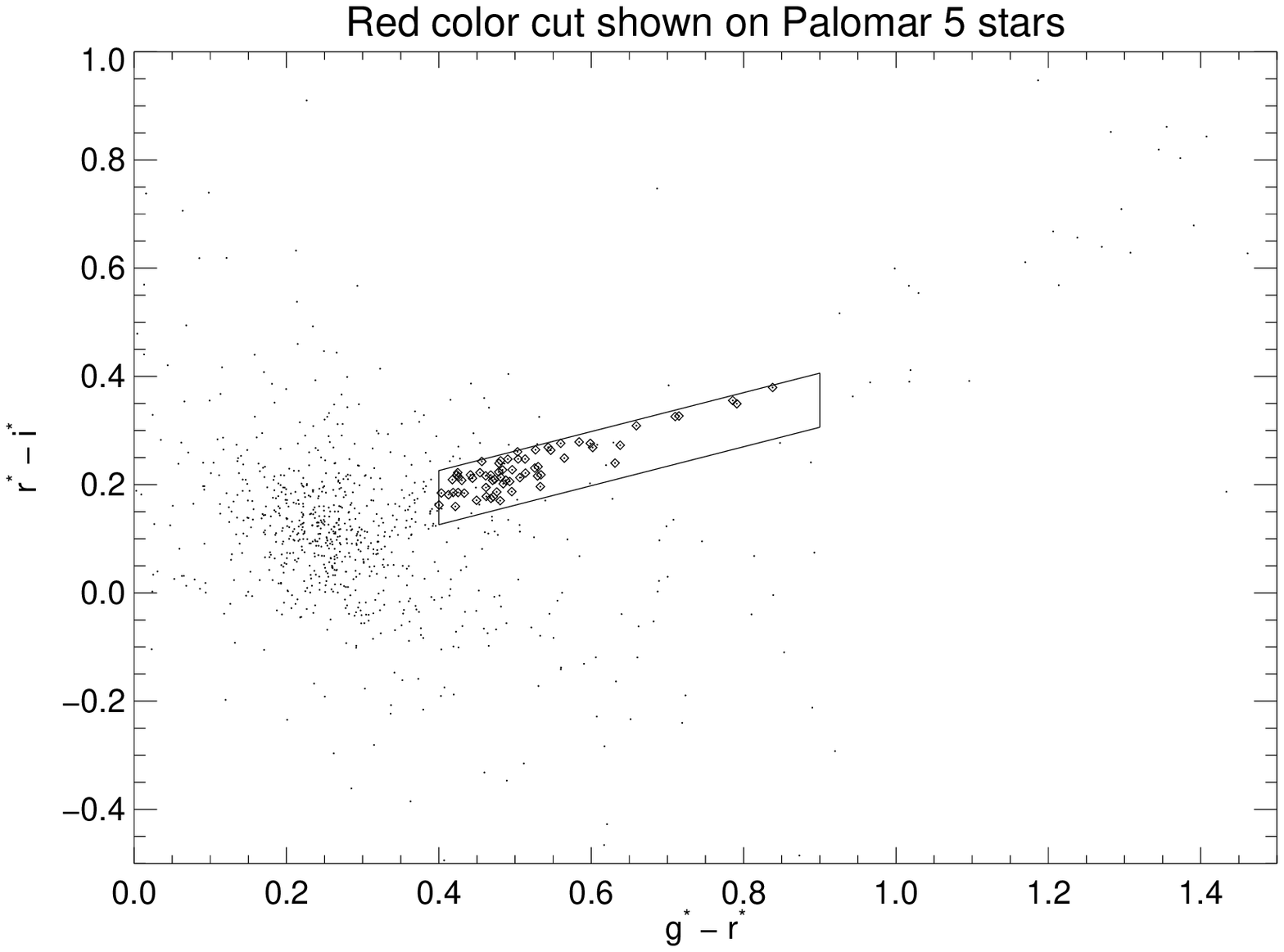}}
%\centerline{\epsfbox{Willman.fig3.eps}}
\ifsubmode
\vskip3.0truecm
\addtocounter{figure}{1}
\centerline{Figure~\thefigure}
\else\figcaption{\figcapGCcut}\fi
\end{figure}
\clearpage
\begin{figure}
\epsfxsize=14.0truecm
\centerline{\epsfbox{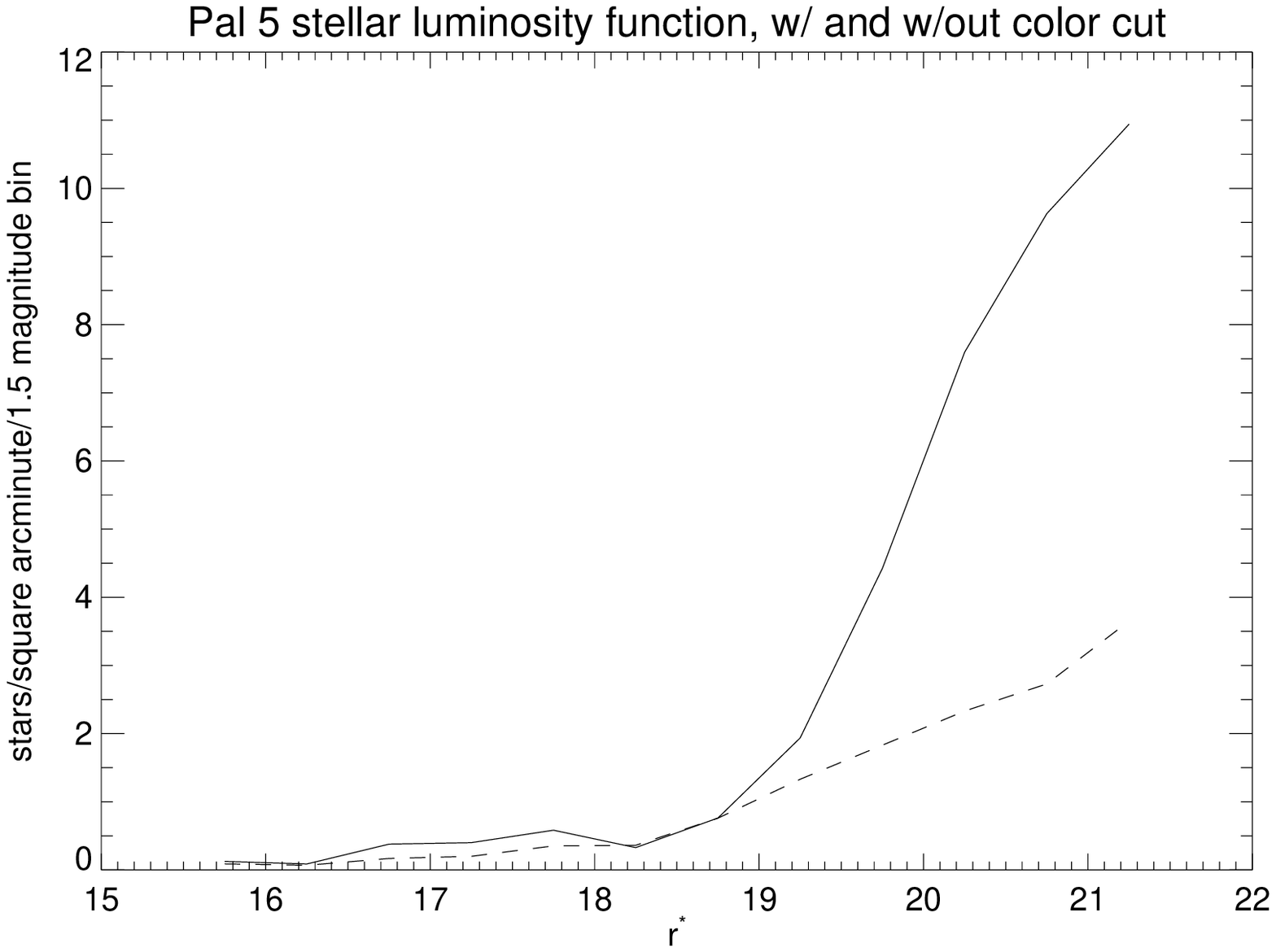}}
%\centerline{\epsfbox{Willman.fig4.eps}}
\ifsubmode
\vskip3.0truecm
\addtocounter{figure}{1}
\centerline{Figure~\thefigure}
\else\figcaption{\figcapGClf}\fi
\end{figure}
\clearpage
\begin{figure}
\epsfxsize=14.0truecm
\centerline{\epsfbox{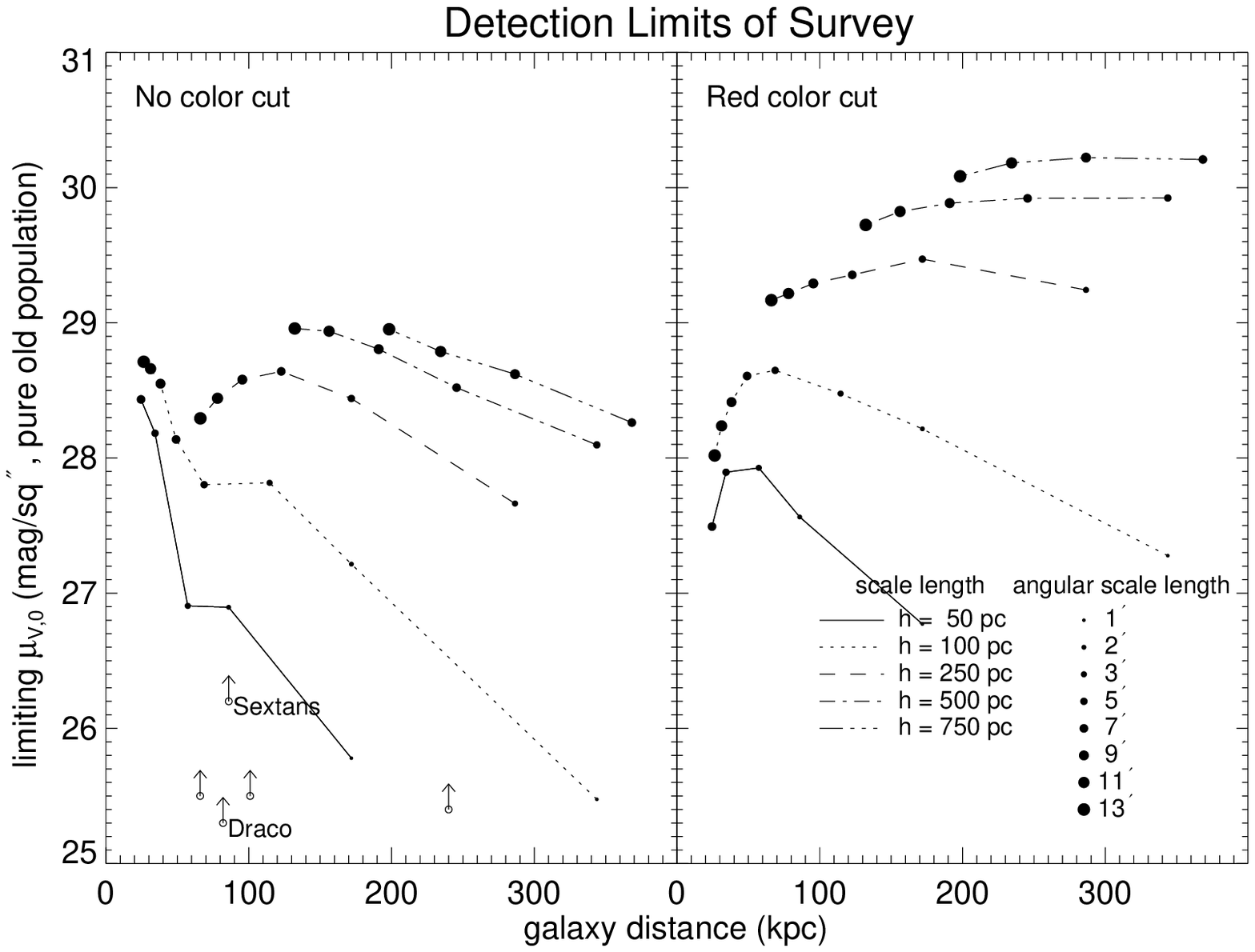}}
%\centerline{\epsfbox{Willman.fig5.eps}}
\ifsubmode
\vskip3.0truecm
\addtocounter{figure}{1}
\centerline{Figure~\thefigure}
\else\figcaption{\figcapfiltersmulin}\fi
\end{figure}
\clearpage
\begin{figure}
\epsfxsize=14.0truecm
\centerline{\epsfbox{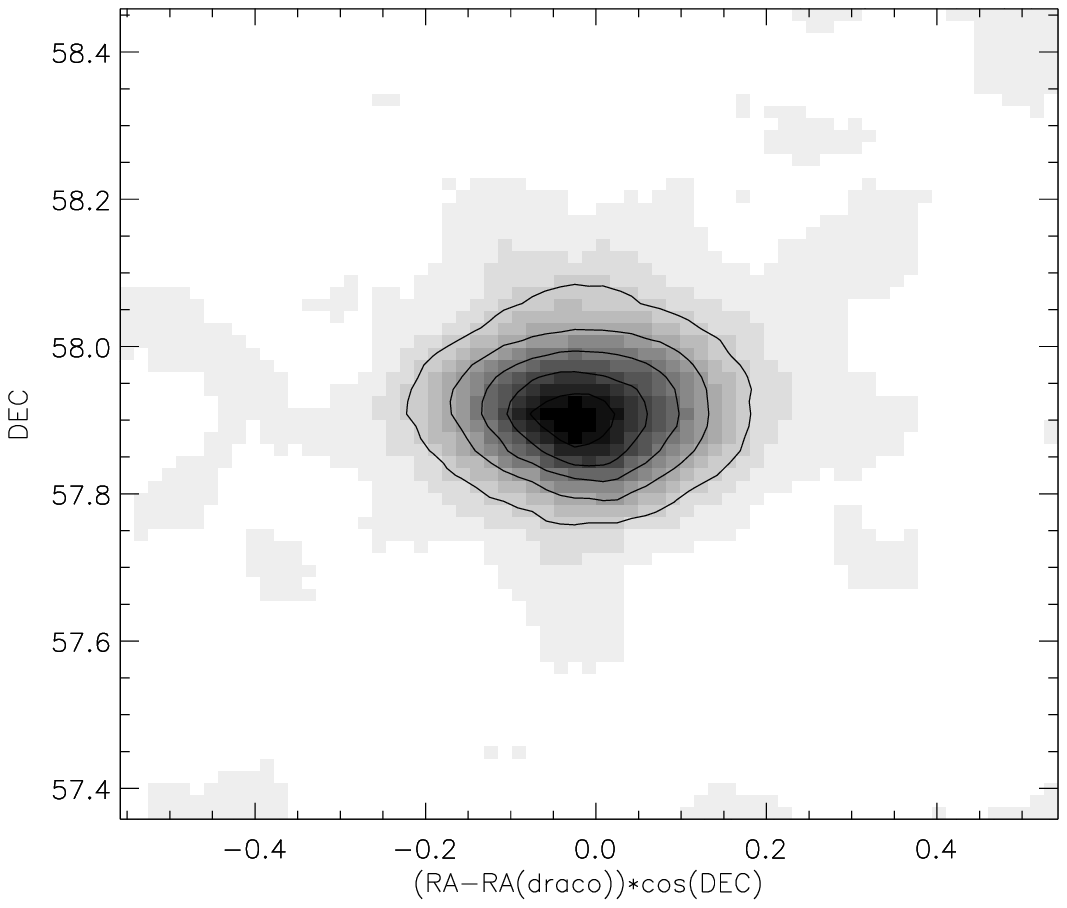}}
%\centerline{\epsfbox{Willman.fig6.eps}}
\ifsubmode
\vskip3.0truecm
\addtocounter{figure}{1}
\centerline{Figure~\thefigure}
\else\figcaption{\figcapdracoimage}\fi
\end{figure}
%%
%%%
%%%

\fi

%%% END OF FIGURES %%%

\clearpage
\ifsubmode\pagestyle{empty}\fi

%% END OF DOCUMENT %%

\end{document}